\documentclass[10pt,twocolumn,letterpaper]{article}

\usepackage{iccv}
\usepackage{times}
\usepackage{epsfig}
\usepackage{graphicx}
\usepackage{amsmath}
\usepackage{amssymb}

\usepackage[para,online,flushleft]{threeparttable}
\usepackage{makecell}
\usepackage{booktabs}
\usepackage{multirow}
\usepackage{bm}
\usepackage{algorithm} 
\usepackage{algorithmic}
\usepackage[algo2e]{algorithm2e} 
\usepackage[pagebackref=true,breaklinks=true,letterpaper=true,colorlinks,bookmarks=false]{hyperref}
\usepackage{adjustbox}
\usepackage{color}
\usepackage[accsupp]{axessibility}  

\usepackage[breaklinks=true,bookmarks=false]{hyperref}

\iccvfinalcopy 

\def\iccvPaperID{10890} 

\ificcvfinal\fi

\begin{document}

\title{Achieving on-Mobile Real-Time Super-Resolution with Neural Architecture and Pruning Search}

\author{Zheng Zhan$^{*1}$, Yifan Gong$^{*1}$, Pu Zhao$^{*1}$, Geng Yuan$^{1}$, Wei Niu$^{2}$, Yushu Wu$^{1}$, Tianyun Zhang$^{3}$,\\
Malith Jayaweera$^{1}$, David Kaeli$^{1}$, Bin Ren$^{2}$, Xue Lin$^{1}$,  Yanzhi Wang$^{1}$\\

$^1$Northeastern University, $^2$College of William \& Mary, $^3$Cleveland State University\\
{\tt\small \{zhan.zhe, gong.yifa, zhao.pu, xue.lin, yanz.wang\}@northeastern.edu,}\\
{\tt\small t.zhang85@csuohio.edu, kaeli@ece.neu.edu, bren@cs.wm.edu}
}

\maketitle
\def\thefootnote{$*$}\footnotetext{Equal contribution}
\ificcvfinal\thispagestyle{empty}\fi

\begin{abstract}
Though recent years have witnessed remarkable progress in single image super-resolution (SISR) tasks with the prosperous development of deep neural networks (DNNs), the deep learning methods are confronted with the computation and memory consumption issues in practice, especially for resource-limited platforms such as mobile devices. To overcome the challenge and facilitate the real-time deployment of SISR tasks on mobile, we combine neural architecture search with pruning search and propose an automatic search framework that derives sparse super-resolution (SR) models with high image quality while satisfying the real-time inference requirement. To decrease the search cost, we leverage the weight sharing strategy by introducing a supernet and decouple the search problem into three stages, including supernet construction, compiler-aware architecture and pruning search, and compiler-aware pruning ratio search. With the proposed framework, we are the first to achieve real-time SR  inference  (with only tens of milliseconds per frame) for implementing 720p resolution with competitive image quality (in terms of PSNR and SSIM) on mobile platforms (Samsung Galaxy S20).  
\end{abstract}


\section{Introduction}

In recent year, people have ever-increasing demands for image processing to achieve higher resolutions, leading to the rapid development of SR. In general, the SR principle is to convert low-resolution images to high-resolution  images with clearer details and more information.  It has been adopted in various applications such as crime scene analysis to identify unnoticeable  evidence or medical image processing for more accurate diagnosis. 

With the fast growth of live streaming and video recording, video contents enjoy high popularity. However, videos often have lower resolution due to the limited communication bandwidth or higher resolution of the display. Besides, live streaming usually has a \emph{real-time}\footnote{We believe targeting sub 100ms can be reasonably called real-time \cite{miller1968response} and we require the real-time implementation to be faster than 50ms.} requirement that the latency of each frame should not exceed a threshold. Thus, it is desirable to achieve real-time SR for video locally.   

Compared with the classic interpolation algorithms to improve image or video resolution, deep learning-based  SR can deliver higher visual qualities by learning the mappings from the low-resolution to high-resolution images from external datasets. 
Despite its superior visual performance, deep learning-based SR  is usually   more expensive  with large amounts of computations and  huge power consumption (typically hundreds of watts on powerful GPUs) \cite{dong2017learning,dong2016accelerating,shi2016real}, leading to difficulties for the real-time implementations.  Moreover, in practice, as SR is often  deployed on edge devices such as mobile phones for live streaming or video capturing due to the wide spread of mobile phones, the limited memory and computing resources on edge devices make it even harder for achieving real-time SR.  

Weight pruning  \cite{wen2016learning,guo2016dynamic,he2019filter} is often adopted to remove the redundancy in DNNs to reduce the resource requirement and accelerate the inference. There are various pruning schemes including unstructured pruning \cite{han2015learning,guo2016dynamic,frankle2018lottery,liu2018rethinking}, coarse-grained structured pruning \cite{min20182pfpce,zhuang2018discrimination,zhu2018ijcai,ma2019tiny,Liu2020Autocompress}, and fine-grained structured pruning \cite{ma2019pconv,dong2020rtmobile,gong2020privacy}.  Unstructured pruning  removes arbitrary weights, leading to irregular pruned weight matrices and limited hardware parallelism. Structured pruning  maintains a full matrix format of the remaining weights such that the pruned model is compatible with GPU acceleration for inference. Recently, fine-grained structured pruning including pattern-based pruning and block-based pruning are proposed to provide a finer pruning granularity for higher accuracy while exhibiting certain regularities which can be optimized with compilers to improve hardware parallelism. To achieve inference acceleration of SR models, we focus on conventional structured pruning and fine-grained structured pruning.

Prior works usually use fixed pruning scheme for the whole model. As different pruning schemes can achieve different SR and acceleration performance, a new optimization dimension is introduced to find the  most-suitable pruning configuration for each layer instead of for the whole model. 
Besides, as the performance of pruning depends on the original unpruned model, it is also essential to search an unpruned starting  model with high SR performance.

In this paper, to facilitate the real-time SR deployment on edge devices, we propose a framework incorporating architecture and pruning search to find the most suitable cell-wise SR block configurations and layer-wise pruning configurations.  Our implementation can achieve real-time SR inference with competitive SR performance on mobile devices.
We summarize our contribution as follows.
\begin{itemize}
 \item We propose an  architecture and pruning search framework to automatically find the best configuration of the SR block in each cell and pruning scheme for each layer, achieving real-time SR implementation on mobile devices with high image quality.
  \item We train a supernet to provide a well-trained unpruned model  for all possible  combinations of the SR block in each supernet cell before the architecture and pruning search. Thus there is no need  to train a separate unpruned model for each combination with multiple epochs, saving tremendous training efforts.  
  \item Different from previous works with fixed pruning scheme for all layers or fixed SR blocks for all cells, we automatically search the best-suited  SR block for each cell and pruning scheme for each layer. 
  To reduce the complexity, we decouple the pruning ratio search and employ Bayesian optimization (BO) to accelerate the SR block and pruning scheme search.
  \item  With the proposed method, we are the first to achieve real-time SR inference (with only tens of milliseconds per frame) for implementing 720p resolution with competitive image quality (in terms of PSNR and SSIM) on mobile platforms (Samsung Galaxy S20). Our achievements facilitate various practical applications with real-time requirements such as live streaming or video communication. 
\end{itemize}





\section{Background and Related Works}
\label{sec:background}
\subsection{Preliminaries on Deep Learning-based SR}


%
SISR aims to generate a high resolution image from the  low-resolution version. The usage of DNNs for SR task was first proposed in SRCNN \cite{dong2014learning} and later works try to improve the upscaling characteristic and image quality with larger networks 
\cite{kim2016accurate,lim2017enhanced,zhang2018residual,zhang2018image,dai2019second}. 
However, SR models are resource-intensive due to  maintaining or upscaling the spatial dimensions of the feature map for each layer. Therefore, the number of multiply-accumulate (MAC) operations is typically counted in gigabits, leading to high inference latency (seconds per image) on a powerful GPU.

Several attempts were made to design lightweight SR models for practical applications, including using upsampling operator at the end of a network \cite{dong2016accelerating,shi2016real}, adopting channel splitting \cite{hui2018fast}, and using wider activation \cite{yu2018wide}. Specifically, work \cite{yu2018wide} proposed WDSR-A and WDSR-B blocks, which are two of the state-of-the-art SR building blocks with high image quality. 
Besides, inspired by the success of neural architecture search (NAS), 
latest SR works try to establish more efficient and lightweight SR models by leveraging  NAS approaches \cite{chu2019fast,song2020efficient,lee2020journey,chu2020multi}. 
But the proposed models are still too large with tremendous resource demands. 
Furthermore, 
they do not consider practical mobile deployments with limited hardware resource. 
For mobile deployment, the winner of the PIRM challenge \cite{vu2018fast} and MobiSR \cite{lee2019mobisr} are the few works that make progress for SR inference on mobiles. But  the latency is still far from real time, requiring nearly one second per frame. 



\subsection{DNN Model Pruning} \label{sec:weight_pruning}
Weight pruning reduces the redundancy in DNNs for less storage and computations.
Existing pruning schemes can be divided into \textit{unstructured pruning}, \textit{coarse-grained structured pruning}, and \textit{fine-grained structured pruning}. 

\begin{figure}[t]
    \centering
    \includegraphics[width=0.75 \columnwidth]{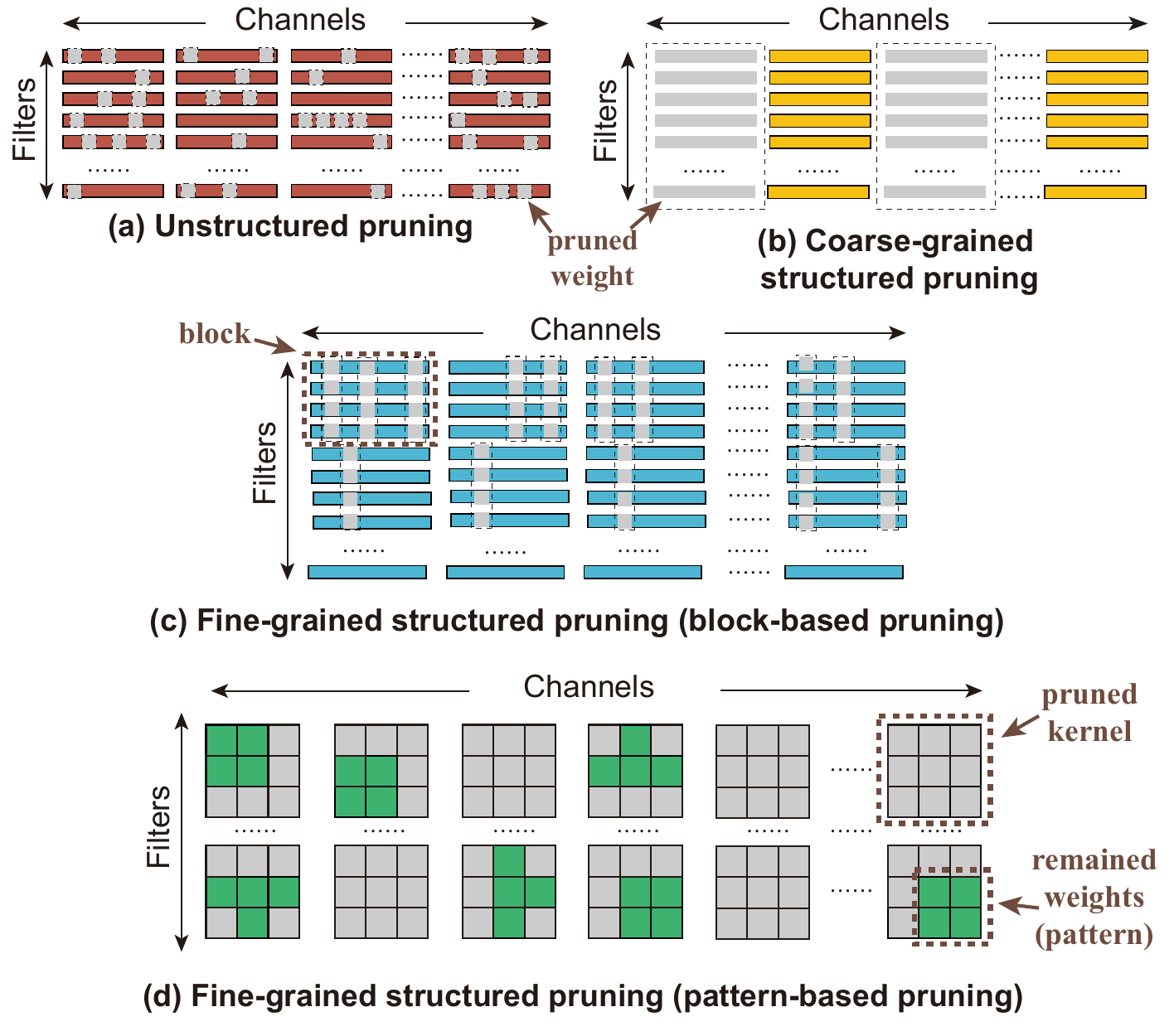}
    \caption{(a) Unstructured pruning; (b) coarse-grained structured pruning (channel); (c) fine-grained structured pruning (block-based); and (d) fine-grained structured sparsity (pattern-based).}
    \label{fig:prune_type}
\end{figure}

Unstructured pruning allows weights at arbitrary locations to be removed~\cite{guo2016dynamic,frankle2018lottery,dai2019nest}, as shown in Figure~\ref{fig:prune_type} (a). Despite the high accuracy, its irregular weight matrices are not compatible with GPU acceleration. 
Coarse-grained structured pruning \cite{wen2016learning,he2017channel,he2019filter,yu2018nisp,hou2020efficient} keeps structured regularity of remaining weights such as channel pruning prunes entire channels as in Figure~\ref{fig:prune_type} (b). 
The key advantage is that a full matrix format is maintained, thus facilitating hardware acceleration. 
However, coarse-grained structured pruning often leads to non-negligible accuracy degradation \cite{Wang2019NonstructuredDW}. 




Fine-grained structured pruning includes block-based pruning \cite{dong2020rtmobile} and pattern-based pruning \cite{niu2020patdnn,ma2019pconv,gong2020privacy}. They incorporate the benefits from fine-grained pruning while maintaining structures that can be exploited for hardware accelerations with the help of compiler. 
Block-based pruning divides the weight matrix of a DNN layer into multiple equal-sized blocks and applies structured pruning independently to each block, as shown in Figure~\ref{fig:prune_type} (c). 
Pattern-based pruning is a combination of kernel pattern pruning and connectivity pruning, as illustrated in Figure~\ref{fig:prune_type} (d). Kernel pattern pruning removes weights by forcing the remaining weights in a kernel to form a specific kernel pattern. 
Connectivity pruning removes entire redundant kernels and is the supplementary to kernel pattern pruning for a higher compression rate. 
With an appropriate pruning regularity degree, compiler-level code generation can be exploited to achieve a high hardware parallelism.

\subsection{DNN Acceleration Frameworks on Mobile}

On-mobile DNN inference has attracted many interests from both industry and academia \cite{lane2016deepx,lane2015deepear,xu2018deepcache,huynh2017deepmon,yao2017deepsense,han2016mcdnn}.
Representative DNN acceleration frameworks, including Tensorflow-Lite \cite{TensorFlow-Lite}, Alibaba MNN \cite{Ali-MNN}, Pytorch-Mobile \cite{Pytorch-Mobile}, and TVM \cite{chen2018tvm}, are designed to support inference acceleration on mobile. Several graph optimization techniques are used in these frameworks, including layer fusion, constant folding, and runtime optimizations on both mobile CPU and GPU.
But the missing piece is that sparse (pruned) models for further speedup are not supported.
Recently, some efforts are made to accelerate pattern-based pruned models on mobile with compiler-based optimizations \cite{niu2020patdnn,ma2019pconv}.
But they suffer difficulties when generalized to DNN layers other than 3$\times$3 convolutional (CONV) layers.


\subsection{Motivation}

State-of-the-art SR methods leverage huge DNNs to pursue high image quality, causing extremely high computation cost. Thus, it is difficult to achieve real-time SR even on powerful GPUs, not to mention mobile devices with limited resource. But due to the widespread of mobile phones and the popular video communication and live streaming applications with high resolution requirements, it is desirable to implement on-mobile real-time SR with high image quality.

\begin{figure}[t]
    \centering
    \includegraphics[width=0.81 \columnwidth]{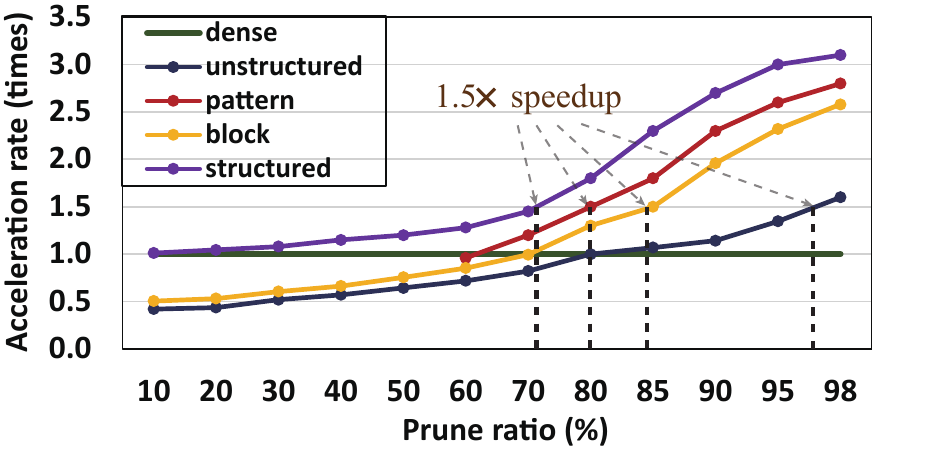}
    \caption{Inference acceleration rate vs. pruning ratio of different pruning schemes. 
    Results are measured on a Samsung Galaxy S20 smartphone, and a typical 3$\times$3 CONV layer in WDSR block with 24/48 input/output channels and 320$\times$180 feature size is used.}
    \label{fig:latency_vs_ratio}
\end{figure}

SR models usually constitute several cascaded SR blocks. Different blocks have different latency performance, while different combinations can form various SR models with different image quality. 
Meanwhile, with weight pruning for acceleration, each layer may prefer a different pruning scheme, resulting in different accuracy and acceleration performance. 
For instance, Figure~\ref{fig:latency_vs_ratio} illustrates the acceleration curves of different pruning schemes on a given 3$\times$3 CONV layer.
Hence, it is desirable to find the best-suited combination of SR blocks and per-layer pruning scheme and ratio to achieve high image quality while satisfying the real-time execution requirement.

Finding the satisfied network architecture and pruning configurations is too complex to be solved manually. Thus an automatic architecture and pruning search method \cite{wang2020apq} is desired.  However, it is expensive to directly search in a large space, including block number (depth), block type, per-layer pruning scheme, and per-layer pruning ratio. Hence, we decouple the search into several stages and solve them separately.

\section{Framework Overview}

The objective is to combine architecture search with pruning search to find sparse SR models facilitating various practical applications such as live streaming or video communication. The sparse SR models should satisfy the real-time inference requirement (with only tens of milliseconds per frame) for high upscaling resolution such as 720p (1280$\times$720) on mobile devices, with competitive image quality with the state-of-the-art methods.

The searching problem involves the determination of the number of stacked cells, the type of selected block in each cell, and pruning scheme and pruning ratio for each layer of the SR network. Direct search in such a high-dimensional search space is computationally expensive. To reduce the search cost in terms of time and computation, we leverage the weight sharing strategy by introducing a supernet and decouple the search problem into three stages: 1) supernet construction, 2) compiler-aware architecture and pruning search, and 3) compiler-aware pruning ratio determination. Supernet construction includes supernet initialization that determines the number of stacked cells, and supernet training that provides a good starting point for the following two steps. Then, a combination of block determination and pruning scheme selection for each layer is performed. The goal is to find a desirable structure that maximizes the image quality while satisfying the target latency $t$ with the aid of compiler optimizations. Specifically, when $t\leq50$ms, the target latency meets the real-time requirement. The following step is the automatic pruning ratio determination with the reweighted dynamic regularization method. We show the overall framework in Figure \ref{fig:overall_flow}.


\section{Supernet Construction}
In architecture and pruning search, the accuracy of a model (architecture) after pruning largely depends on the accuracy of unpruned starting model. To obtain the well-trained starting models for various architectures with satisfying SR performance, the straightforward method is to perform training for each new architecture, which usually costs huge training efforts.  
Instead of training separate models respectively, we  train a supernet such that, for any new model, we can activate the corresponding path in the supernet to derive the well-trained unpruned model immediately without further efforts to train each new model from scratch.
Thus, the supernet can significantly reduce the training time for the unpruned models, 
thereby accelerating the search.

\begin{figure*} [t]
     \centering
     \includegraphics[width=0.857\textwidth]{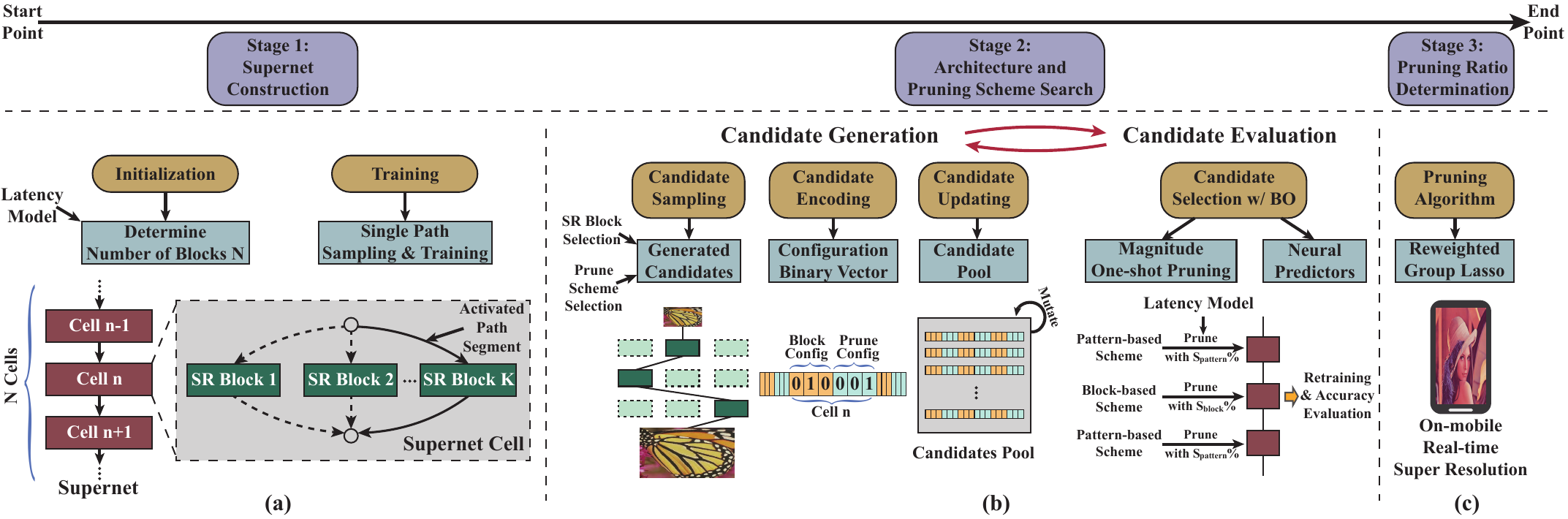}  
     \caption{Framework overview. The framework is composed of three stages to reduce the search cost: (a) stage 1: supernet construction, (b) stage 2: architecture and pruning search, and (c) stage 3: pruning ratio determination. }
    \label{fig:overall_flow}  
\end{figure*}

The architecture of the supernet is illustrated in the Figure \ref{fig:overall_flow} (a).
We encode the architecture search space $\mathcal{A}$ with a supernet, denoted as $\mathcal{S}(\mathcal{A},W)$, where $W$ represents the weight collection. 
The supernet is constituted of $N$ stacked cells and each cell contains $K$ SR block choices. In our work, we adopt WDSR-A and WDSR-B, which are two highly efficient SR blocks with high image quality, as block choices. Note that our framework is not restricted by the WDSR blocks and can be generalized to different kinds of 
SR residual blocks. The output of each SR block $k$ in cell $n$ connects with all of the SR blocks in the next cell $n+1$. 
We define the choice of one SR block (WDSR-A or WDSR-B)  for each supernet cell as one path segment, and all of the possible combinations of the $N$ path segments form the architecture search space $\mathcal{A}$ with a size of $K^N$. Then one path is the collection of $N$ path segments for all cells, denoting one SR candidate model.
During supernet computation, only one path is activated while other unselected SR blocks do not participate into the computation.



To construct a supernet, there are two necessary steps: 1) determine  the number of stacked cells of the supernet and initialize the supernet, and 2) fully train the supernet to provide a good starting point with high image quality and low overhead for the following SR candidate nets search.

\subsection{Determine Cell Number with Latency Models} \label{sec: latency_model}


The number of stacked cells $N$ of the supernet should be determined beforehand to guarantee the SR candidate models have the potential to satisfy the target latency $t$ on mobile devices. Several widely used techniques in SR such as pixel shuffling (a.k.a., sub-pixel convolution) and global residual path are often hard to optimize and accelerate, resulting in a fixed latency overhead. 
Moreover, the skip (identity) connection structure in a block of a cell leads to a certain execution overhead that is difficult to be reduced and is accumulated with the number of stacked blocks.


To determine the number of stacked cells, we build a latency model enabling fast and precise estimation of the overall model inference latency on the target device (e.g., Samsung S20 smartphone).
The latency model contains the look-up-tables of inference latency for different types of layers used in SR models (e.g., 1$\times$1 CONV, 3$\times$3 CONV, 5$\times$5 CONV, skip connection, and  sub-pixel convolution).
For each layer type, several different settings are considered, including the number of filters and input and output feature map size.
Our latency model is compiler-aware, built by measuring real-world inference latency on the target device with compiler optimizations incorporated. More details about our compiler optimization techniques are shown in Appendix \ref{app: compiler}.
The latency model building time can be ignored since no training process is involved, and the building process can be conducted in parallel with the supernet training. We only build once for a specific device.
Moreover, we also include the sparse inference latency for different types of layers under different pruning schemes and pruning ratios in our latency model, which will be used in the pruning search stage (more discussion in Section~\ref{sec:candidates_generation}).

Therefore, the overall inference latency on the target device can be estimated by accumulating the per-layer latency inquired from our latency model.
With a target latency $t$ for the SR candidate models, the suitable number of stacked cells can be determined. Furthermore, decoupling the supernet depth determination from the search space of the candidate SR models can greatly reduce the search complexity.

\subsection{Supernet Training}

After the supernet is initialized, the next step is to train its weights $\mathcal W$ to minimize the loss function $\mathcal{L}(\mathcal{A},\mathcal W)$. 
The well-trained supernet  provides a good starting point for the following network architecture and pruning search as candidate net architecture $a$ directly inherits weights from the path  $\mathcal W(a)$ in the supernet. Note that the weights $\mathcal W$ of the supernet should be optimized in a way that all the candidate architectures $a \in \mathcal{A}$ with weights $\mathcal W(a)$ are optimized simultaneously. 
However, jointly optimizing the architecture parameters $a$ and model parameters $\mathcal W(a)$  often introduces extra complexities. Furthermore, it may lead to the situation that some nodes in the graph are well trained while others are poorly trained, incurring unfair comparison for paths of different levels of maturity in the supernet.

To mitigate this problem, we adopt a single-path sampling \& training strategy to accelerate the convergence of supernet training. Specifically, for each training batch, we only activate and train one random path while other unselected SR blocks are skipped. 
In this way, the architecture selection and model weights updating are  decoupled. This strategy is hyper-parameter free, and each path is a SR model providing a well-trained unpruned starting point for the following architecture and pruning search.




\section{Architecture and Pruning Search}
We define each \emph{architecture and pruning candidate} as a configuration to select one SR block for each supernet cell together with choosing the pruning scheme for each layer.
The architecture and pruning search aims to find the best cell-wise SR block selection and layer-wise pruning scheme configuration, i.e., the candidate with the highest image quality satisfying the target latency $t$. 
The search consists of two main steps: 1)  candidate generation and 2)  candidate evaluation. 
In each iteration, candidate generation samples architecture and pruning candidates, which are further evaluated in the candidate evaluation process. 
To improve search efficiency, we adopt evolutionary-based candidate updating in candidate generation and BO in candidate evaluation to obtain the best candidate. 


\subsection{Candidate Generation}
\label{sec:candidates_generation}
\subsubsection{Candidate Sampling}
\label{sec:candidates_sampling}
The  candidate generation samples \emph{architecture and pruning candidates} from the search space. 
Each candidate $g$ is a directed acyclic graph denoting the cell-wise SR block selection and  layer-wise pruning scheme selection.
For SR block selection in each supernet cell, we can choose from WDSR-A block or WDSR-B block. 
For the pruning scheme, 
we can choose  channel pruning \cite{wen2016learning}, pattern-based pruning \cite{ma2019pconv}, or block-based pruning \cite{dong2020rtmobile} for each layer. 
Different from previous works with fixed pruning scheme  for all layers, we can  choose different pruning schemes for different layers, which is also supported by our compiler code generation. Note that the difference between the candidate $g$ and the candidate network architecture $a$ is that $g$ includes the per-layer pruning scheme selection.

We encode each candidate with a binary vector by assigning a binary feature for each possible cell-wise block choice and layer-wise pruning scheme selection, denoting whether the block or pruning scheme is adopted or not.

\paragraph{Decoupling pruning ratio search.} 
To prune the model, we also need to configure the layer-wise pruning ratio corresponding to each pruning scheme. As it is  expensive to search the continuous pruning ratio values for each layer, at this step, we simply set the layer-wise pruning ratio to a minimal value satisfying the target latency $t$. Therefore, we can focus on pruning scheme search first. 
To determine the minimal pruning ratio, 
we can estimate the latency of the unpruned model $t'$ and the target latency $t$, and obtain the minimal speedup required for the whole model, which is $t'/t$.
To satisfy the overall speedup, we simply require each layer to achieve this minimal speedup $t'/t$\footnote{We  find that each layer in the SR model has similar computation amount. Thus it is reasonable to adopt the same speedup for each layer.}. Then, according to the latency model (detailed in Section \ref{sec: latency_model}) and the layer-wise speedup, we can obtain the layer-wise minimal pruning ratio corresponding to each pruning scheme.

\subsubsection{Candidate Updating}
In each iteration, we need to generate a pool of new  candidates. To make the candidates updating more efficient, the evolutionary-based candidate updating method is adopted. We keep a record of  all evaluated  candidates with their evaluation performance. To generate new  candidates, we  mutate the candidates  with the best evaluation performance  in the records by randomly changing one SR block of one random cell or one pruning scheme of one  random layer. Specifically, we first select $H$ candidates with  highest evaluation performance, and mutate each of them iteratively until  $C$ new proposals are derived.

\subsection{Candidate Evaluation} 

As it incurs a high time cost to prune and retrain the model following each candidate,  BO  \cite{chen2018bayesian} is adopted to expedite the candidate evaluation. 
With the generated $C$ candidates, we first use BO to select  $B$ ($B<C$)  candidates with  potentially better performance. Then the selected candidates are evaluated to obtain the accurate SR performance while the unselected candidates are not evaluated. 
The number of actually evaluated candidates is reduced in this way. 

BO includes two main components, i.e., training an ensemble of neural predictors and selecting candidates based on acquisition function values enabled by the predictor ensemble. 
To make use of BO, the ensemble of  neural predictors provides an average SR prediction with its corresponding  uncertainty estimate for each unseen candidate. Then BO is able to choose the candidate which maximizes the acquisition function. 
We show the full algorithm in Algorithm \ref{alg: full} and specify BO in the following.

\begin{algorithm}[tb]
        \small
	    \caption{Evaluation with  BO }\label{alg: full}
	\begin{algorithmic}
		\STATE {\bf Input:}  Observation data $\mathcal{D}$, BO batch size $B$, BO acquisition function $\phi(\cdot)$
		\STATE {\bfseries Output:} The best candidate $g*$
        \FOR{steps}
        		\STATE Generate a pool of candidates $ \mathcal{G}_c $;
        		\STATE Train an ensemble of neural predictors with $\mathcal{D}$;
        		\STATE Select $\{ \hat{g}^{ i} \}_{i=1}^B = \arg\max_{g \in  \mathcal{G}_c } \phi (g)$;
        		\STATE Evaluate the candidate and obtain reward  $\{r^{ i}\}_{i=1}^B$ of $\{ \hat{g}^{ i} \}_{i=1}^B$;
        		\STATE $\mathcal{D}\leftarrow \mathcal{D} \cup (\{ \hat{g}^{  i} \}_{i=1}^B , \{r^{i}\}_{i=1}^B)$;
    	\ENDFOR
	\end{algorithmic}
\end{algorithm}

\subsubsection{Bayesian Optimization with Neural Predictors}

\paragraph{Neural predictor.}
The neural predictor is a  neural network repeatedly trained on the current set of evaluated candidates with their evaluation performance  to  predict the reward  of unseen candidates. 
It is a neural network with  8 sequential fully-connected layers of  width 40 trained by the Adam optimizer with a learning rate of 0.01. 
For the loss function to train neural predictors, mean absolute percentage error (MAPE)  is adopted as it can give a higher weight to candidates with higher evaluation performance:
{\small
\begin{equation}
\mathcal{L}(m_\text{pred}, m_\text{true}) =
    \frac{1}{n}\sum_{i=1}^n \left| \frac{m_\text{pred}^{(i)} - m_\text{UB}}{m_\text{true}^{(i)} - m_\text{UB}} - 1\right|,\label{eq:mape}
\end{equation}}%
where $m_\text{pred}^{(i)}$ and $m_\text{true}^{(i)}$ are the predicted and true values, respectively, of the reward for the $i$-th candidate in a batch, and $m_\text{UB}$ is a global upper bound on the maximum true reward.  Note that the training of the predictors does not cost too much efforts  due to their simple architectures.

\paragraph{Ensemble of  neural predictors.} 
To incorporate BO, it also needs  an uncertainty estimate for the prediction. So we adopt an ensemble of neural predictors  to provide the uncertainty estimate. More specifically, we train $P$ neural predictors using different random weight initializations and training data orders. Then for any candidate $g$, we can obtain the mean and standard deviation of these $P$ predictions. Formally,  we  train an ensemble of $P$ predictors,
$\{f_p\}_{p=1}^P$, where $f_p(g)$ provides a predicted reward for a candidate $g$.
The mean prediction and its deviation are given by
{\small
\begin{equation}
\hat{f}(g) = \frac{1}{P}\sum_{p=1}^P f_p(g), \ \text{and} \ \ \hat{\sigma}(g)  = \sqrt{\frac{\sum_{p=1}^P (f_p(g) - \hat{f}(g))^2}{P-1}}. 
\end{equation}}%

\paragraph{Selection with acquisition function.}
After training an ensemble of neural predictors, we can obtain the acquisition function value for candidates in the pool and select a small portion of candidates with the largest  acquisition function values. We choose the  upper confidence bound (UCB) \cite{srinivas2010gaussian} as the acquisition function shown below:  
{\small
\begin{align}    
    \phi_\text{UCB}(g) = \hat{f}(g) + \beta \hat{\sigma}(g),
\end{align}}%
where the tradeoff parameter   $\beta $ is set to 0.5.

\subsubsection{Evaluation with magnitude-based pruning}

After selecting the candidates from the pool, we need to measure the performance of the selected candidate $g$ to update the neural predictors. For faster evaluation, magnitude-based pruning framework \cite{han2015learning} (with two steps including pruning and retraining) is adopted to perform the actual pruning for candidate $g$ to obtain its evaluation performance. Note that multiple candidates can be  evaluated  in parallel. Once the evaluation finishes, their actual performances are recorded as a reference such that the candidate generation can sample better candidates.

\section{Pruning Ratio Determination}
\label{sec: ratio}

After finding the best SR block configuration for each cell and the pruning scheme for each layer, we adopt a pruning ratio determination process to derive the suitable layer-wise pruning ratio.
Unlike prior works (i.e., group Lasso regularization~\cite{wen2016learning,he2017channel,liu2017learning} or Alternating Direction Methods of Multipliers (ADMM)~\cite{zhang2018systematic,ren2019admm,li2019compressing}) that suffers from significant accuracy loss or complicated pruning ratio tuning, we adopt the reweighted group Lasso~\cite{candes2008enhancing,ma2020blk} method to determine the layer-wise prune ratio automatically.

The basic idea is to assign a penalty to each weight or pruning pattern, and dynamically reweight the penalties. 
More specifically, during the training (pruning) process, the reweighted method reduces the penalties on weights with larger magnitudes, thus enlarging the more critical weights, 
and increases the penalties on weights with smaller magnitudes, thus decreasing negligible weights. 
After convergence, the desired pruning ratio for each layer is determined automatically.
The reweighted method can be adopted for different pruning schemes and layer types. We show the detailed reweighted pruning algorithm in Appendix \ref{app: reweightedl1}.

\section{Experiments}


\subsection{Methodology}

\noindent\textbf{Datasets:} 
All SR models were trained on the training set of the DIV2K \cite{Agustsson_2017_CVPR_Workshops} dataset with 800 training images. For the evaluation, four benchmark datasets 
Set5 \cite{bevilacqua2012low}, Set14 \cite{yang2010image}, B100 \cite{martin2001database} and Urban100 \cite{huang2015single} are employed as test sets, and the PSNR and SSIM indices are calculated on the luminance channel (a.k.a. Y channel) of YCbCr color space.

\noindent\textbf{Evaluation Platforms and Running Configurations:} 
The training codes are implemented with the PyTorch API. 8 Nvidia TITAN RTX GPUs are used to conduct the architecture and pruning search. We train an ensemble of 20 predictors and 8 models are evaluated in parallel in each step. Since we start from a well-trained 
supernet, we retrain 2 epochs for each one-shot pruned candidate model for fast evaluation. The search process takes 6 GPU days. The latency is measured on the GPU of an off-the-shelf Samsung Galaxy S20 smartphone, which has the Qualcomm Snapdragon 865 mobile platform with a Qualcomm Kryo 585 Octa-core CPU and a Qualcomm Adreno 650 GPU. Each test takes 50 runs on different inputs with 8 threads on CPU, and all pipelines on GPU. As different runs do not vary greatly, only the average time is reported for readability. 


\subsection{Comparison with State-of-the-Art}

\begin{figure}[t]
     \centering
     \includegraphics[width=0.9\columnwidth]{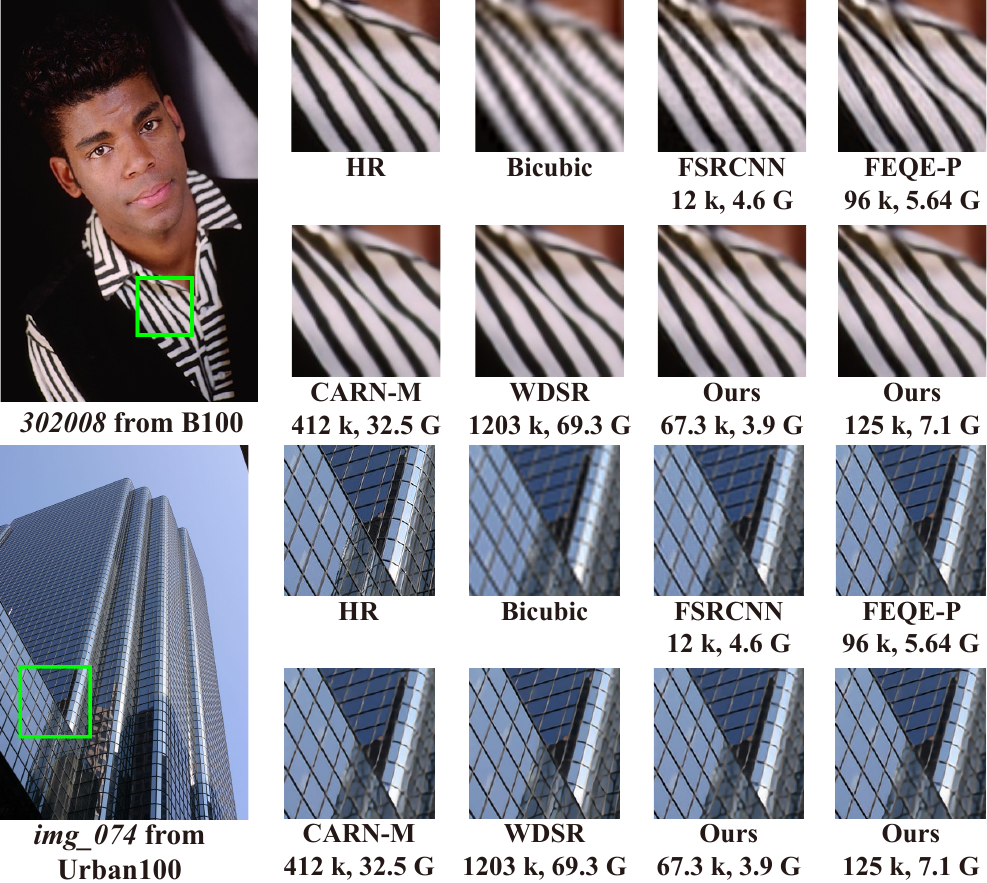}  
     \caption{Visual comparison with other SR models on $\times$4 scale. Model parameters and MACs are listed under model name. More results can be found in Appendix \ref{sec:visual_compare}.}
    \label{fig:patch}  
\end{figure}

The comparison of our SR model obtained through the proposed framework with state-of-the-art methods are shown in Table \ref{table:result_sr}. Some extremely large models \cite{zhang2018residual,zhang2018image,dai2019second} 
could take several seconds for them to upscale only one image on a large GPU. Therefore, those results are not included in Table \ref{table:result_sr}. PSNR and SSIM are adopted as metrics to evaluate the image quality by convention. The evaluations are conducted on tasks with different scales including $\times$2, $\times$3, and $\times$4.  For a fair comparison, we start from different low resolution inputs but the outputs have the same high resolution (720p--$1280\times  720$).


\begin{table*}[t]
\scriptsize\sffamily
    \centering
    \renewcommand{\arraystretch}{1.2}
\begin{adjustbox}{max width=1\textwidth}
\begin{tabular}{@{} l l *{2}{r} *{4}{l} @{}}
\toprule
\bfseries Scale
& \bfseries \begin{tabular}{@{}l@{}}Model\end{tabular}
& \bfseries \begin{tabular}{@{}r@{}}Params \\(K)\end{tabular}
& \bfseries \begin{tabular}{@{}r@{}}Multi-Adds \\(G)\end{tabular}
& \bfseries \begin{tabular}{@{}r@{}}Set5\\(PSNR/SSIM)\end{tabular}
& \bfseries \begin{tabular}{@{}r@{}}Set14\\(PSNR/SSIM)\end{tabular}
& \bfseries \begin{tabular}{@{}r@{}}B100\\(PSNR/SSIM)\end{tabular}
& \bfseries \begin{tabular}{@{}r@{}}Urban100\\(PSNR/SSIM)\end{tabular}
\\[2ex]
    \toprule
    \multirow{13}{*}{\rotatebox[origin=c]{0}{\footnotesize $\times$ 2}} 
    & \textsc{SRCNN}~\cite{dong2014learning} &  57 & 52.7 & 36.66/0.9542 & 32.42/0.9063 & 31.36/0.8879 & 29.50/0.8946 \\
    & \textsc{FSRCNN}~\cite{dong2016accelerating} & 12 & 6.0 & 37.00/0.9558 &32.63/0.9088 &31.53/0.8920 &29.88/0.9020 \\
    & \textsc{MOREMNAS-C}~\cite{chu2020multi} & 25 & 5.5 & 37.06/0.9561 & 32.75/0.9094 & 31.50/0.8904 &  29.92/0.9023 \\
    & \textsc{TPSR-NOGAN}~\cite{lee2020journey} & 60 &14.0 &37.38/0.9583 &33.00/0.9123 &31.75/0.8942 &30.61/0.9119\\
    & \textsc{LapSRN}~\cite{lai2017deep} & 813& 29.9& 37.52/0.9590 &33.08/0.9130 &31.80/0.8950 &30.41/0.9100\\
    & \textsc{CARN-M}~\cite{hui2018fast} & 412 &91.2& 37.53/0.9583 &33.26/0.9141 &31.92/0.8960 &31.23/0.9193\\
    & \textsc{FALSR-C}~\cite{chu2019fast} & 408 &93.7 &37.66/0.9586 &33.26/0.9140 &31.96/0.8965 &31.24/0.9187 \\
    & {\textsc{ESRN-V}}~\cite{song2020efficient} & 324 & 73.4 & 37.85/0.9600 &33.42/0.9161 &32.10/0.8987 &31.79/0.9248 \\
    & \textsc{EDSR}~\cite{lim2017enhanced} & 1518 & 458.0 & 37.99/0.9604 & 33.57/0.9175 & 32.16/0.8994 &31.98/0.9272 \\
    & \textsc{WDSR}~\cite{yu2018wide} & 1203 & 274.1 & 38.10/0.9608 & 33.72/0.9182 & 32.25/0.9004 &32.37/0.9302 \\
    & \textbf{Ours ($t=450$}ms\textbf{)}  & 106 & 24.3 & 37.81/0.9599 & 33.37/0.9153 & 32.07/0.8980 & 31.58/0.9225 \\
    & \textbf{Ours ($t=150$}ms\textbf{)} & 52 & 11.7 & 37.52/0.9582 & 33.24/0.9140 & 31.88/0.8953 & 31.18/0.9180 \\
    & \textbf{Ours ($t=50$}ms,\textbf{real-time)}  & 14 & 3.1 & 37.32/0.9549 & 33.17/0.9071 &  31.67/0.8885 & 30.35/0.8986 \\
    \midrule
    \multirow{14}{*}{\rotatebox[origin=c]{0}{\footnotesize $\times$ 4}}
    
    & \textsc{SRCNN}~\cite{dong2014learning} & 57 & 52.7 & 30.48/0.8628 &27.49/0.7503 &26.90/0.7101 &24.52/0.7221 \\
    & \textsc{FSRCNN}~\cite{dong2016accelerating} & 12 & 4.6& 30.71/0.8657& 27.59/0.7535 &26.98/0.7150 &24.62/0.7280 \\
    & {\textsc{TPSR-NOGAN}}~\cite{lee2020journey} & 61 &3.6 &31.10/0.8779 &27.95/0.7663 &27.15/0.7214 &24.97/0.7456 \\
    & {\textsc{FEQE-P}}~\cite{vu2018fast} & 96 & 5.6 &31.53/0.8824 & 28.21/0.7714 & 27.32/0.7273 & 25.32/0.7583 \\
    & {\textsc{DI-based}}~\cite{hou2020efficient} & 92 & 7.0& 31.84/0.889 &28.38/0.775 &27.40/0.730 &25.51/0.765 \\
    & \textsc{CARN-M}~\cite{hui2018fast} & 412 &32.5& 31.92/0.8903 &28.42/0.7762 &27.44/0.7304 &25.62/0.7694\\
    & {\textsc{ESRN-V}}~\cite{song2020efficient} & 324 & 20.7 & 31.99/0.8919 &28.49/0.7779 &27.50/0.7331 &25.87/0.7782 \\
    & \textsc{EDSR}~\cite{lim2017enhanced} & 1518 & 114.5 & 32.09/0.8938 &28.58/0.7813 &27.57/0.7357 &26.04/0.7849 \\
    &{\textsc{DHP-20}~\cite{li2020dhp}}& 790 & 34.1 & 31.94/\quad--- & 28.42/\quad--- & 27.47/\quad--- & 25.69/\quad--- \\ 
    &{\textsc{IMDN}~\cite{hui2019lightweight}}& 715 & --- & 32.21/0.8948 &28.58/0.7811 & 27.56/0.7353 & 26.04/0.7838\\ 
    & {\textsc{WDSR}}~\cite{yu2018wide} & 1203 & 69.3 & 32.27/0.8963 &28.67/0.7838 &27.64/0.7383 &26.26/0.7911 \\
    & \textbf{Ours ($t=170$}ms\textbf{)} & 125 & 7.1 & 31.93/0.8906 & 28.42/0.7763 & 27.44/0.7307 & 25.66/0.7715 \\
    & \textbf{Ours ($t=120$}ms\textbf{)} & 67 & 3.9 & 31.77/0.8886 & 28.34/0.7730 & 27.33/0.7280 & 25.41/0.7615 \\
    & \textbf{Ours ($t=40$}ms,\textbf{ real-time)} & 12 & 0.7 & 30.74/0.8671 & 27.68/0.7562 & 26.98/0.7156 & 24.65/0.7299 \\
    \bottomrule
    \multicolumn{3}{l}{$\dag$ Results on $\times$3 scaling task are shown in Appendix \ref{sec:3upscaling_result}. 
    }  
\end{tabular}
\end{adjustbox}
    \caption{Comparison of searched results with state-of-the-art efficient SR models.}
        \label{table:result_sr}
\end{table*}

To make a comprehensive study, we set the target latency $t$ to different values for each scale. Particularly, as real-time execution typically requires at least 20 frames per second (FPS), we adopt $t=50$ms for $\times2$ and $\times$3 upscaling task and $t=40$ms for $\times4$ upscaling task to obtain models that satisfy real-time inference requirement.
As shown in Table \ref{table:result_sr}, with a target latency $t=450$ms, our model outperforms CARN-M and FALSR-C with higher PSNR/SSIM using much fewer MACs for a $\times2$ upscaling. With $t=150$ms, our model has better PSNR/SSIM than FSRCNN, MOREMNAS-C, and TPSR-NOGAN with similar or even fewer MACs. Furthermore, both of our models for the two different target latency cases achieve higher PSNR/SSIM with fewer MACs compared with SRCNN and LapSRN. Compared with ESRN-V, EDSR, and WDSR, our model greatly saves the MACs while still maintaining high PSNR/SSIM. Specially, we even obtain a extremely lightweight model that meets the real-time requirement by setting $t=50$ms and the model still maintains satisfying PSNR/SSIM.
As for the $\times$4 scaling task, our model obtained with a target latency $t=120$ms prevails SRCNN, FSRCNN and FEQE-P over MACs, PSNR, and SSIM on the four datasets. With a target latency $t=170$ms, our model outperforms DI-based and CARN-M in PSNR/SSIM with similar or even much fewer MACs. Moreover, with $t=40$ms, our model attains real-time inference while keeping competitive PSNR/SSIM.

\subsection{Searched Results for Real-Time SR on Mobile}

\begin{figure}[t]
     \centering
     \includegraphics[width=0.9\columnwidth]{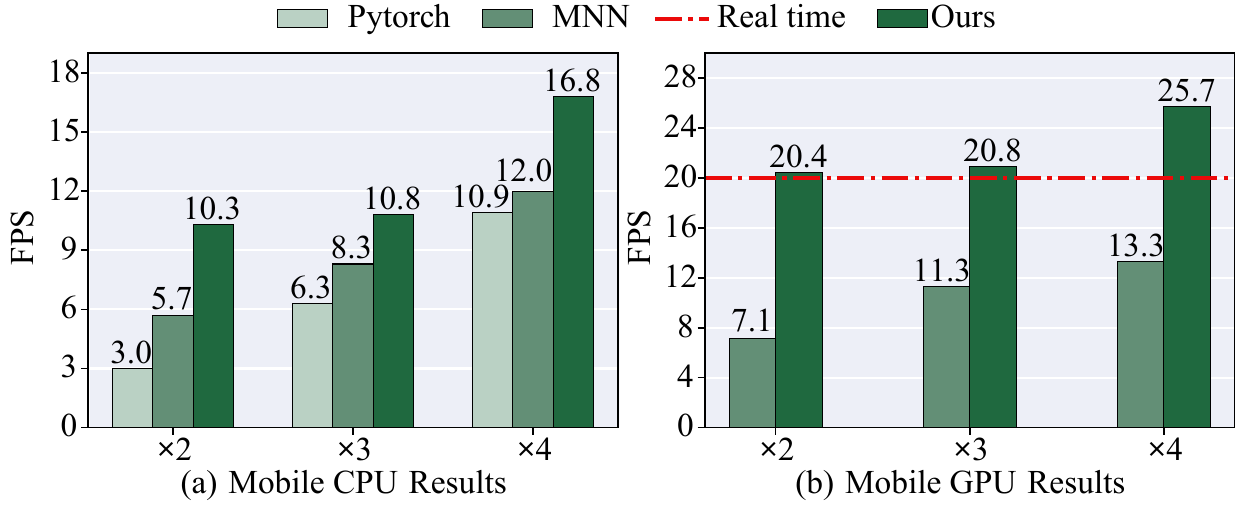}  
     \caption{On-mobile inference comparisons with state-of-the-art mobile acceleration frameworks.}
    \label{fig:fps_sr}  
\end{figure}

We further examine the real-time performance of our SR model assisted with the compiler-based optimizations. As shown in Figure \ref{fig:fps_sr}, with the same SR model derived with the proposed method,   
our method with compiler optimizations achieves the highest FPS for various scales compared with implementations by other acceleration frameworks including MNN \cite{Ali-MNN} and PyTorch Mobile \cite{Pytorch-Mobile}. 
The models are obtained by setting $t=50$ms for $\times 2$ and $\times 3$, and $t=40$ms for  $\times 4$. We can observe from Figure \ref{fig:fps_sr} that our proposed method can satisfy the real-time requirement with a FPS higher than 20 for $\times 2$ and $\times 3$, and higher than 25 for $\times 4$. 

MobiSR and FEQE-P also conduct SR inference on mobile devices. They achieve $2792$ms and $912$ms inference latency on a mobile GPU, respectively, which are far from the real-time requirement.
We highlight that we are the first to achieve real-time SR inference (higher than 20 FPS for  $\times 2$ and $\times 3$, and  25 for $\times 4$) for implementing 720p resolution upsaling with competitive image quality (in terms of PSNR) on mobile platforms (Samsung  Galaxy  S20).


\subsection{Ablation study}

We investigate the influence of architecture search and pruning search separately. For $\times 2$ upscaling, architecture search only achieves a 37.84 PSNR on Set5, slightly higher than ours. But as the computations are not reduced by pruning, it suffers from low inference speed (1.82 FPS). Starting from WDSR blocks, pruning search only with $t=150$ms achieves 6.8 FPS with a lower PSNR (37.40 on Set5). Thus, we can see that pruning search significantly improves the speed performance while architecture search helps mitigate the SR performance loss due to pruning.

To promote the reproducibility  and evaluate speedup using the same framework, we  also implement our derived models and other baseline models including  CARN-M \cite{hui2018fast} and FSRCNN \cite{dong2016accelerating} with the open-source MNN framework. We compare their PSNR and FPS performance and observe that we can achieve higher FPS and PSNR  than the baselines. More details are attached in Appendix \ref{app: mnn}.

\section{Conclusion}
We combine architecture search with pruning search and propose an automatic search framework that derives sparse SR models satisfying real-time execution requirement on mobile devices with competitive image quality.

\section*{Acknowledgment}

{The work is partly supported by Army  Research  Office/Army  Research  Laboratory  via  grant  W911NF-20-1-0167 (YIP) to  Northeastern University, the NSF  CCF-2047516 (CAREER), and Jeffress Memorial Trust Awards in Interdisciplinary Research to William \& Mary.}

{\small
\bibliographystyle{ieee_fullname}
\bibliography{main}
}

\clearpage

\appendix

\section*{Appendix}
\setcounter{equation}{0}
\setcounter{figure}{0}
\setcounter{table}{0}
\makeatletter
\renewcommand{\theequation}{A\arabic{equation}}
\renewcommand{\thefigure}{A\arabic{figure}}
\renewcommand{\thetable}{A\arabic{table}}
\appendix

\section{Compiler Optimization Details} \label{app: compiler}
We provide more details of our compiler optimizations in this section. Different from prior DNN inference acceleration frameworks \cite{TensorFlow-Lite,Pytorch-Mobile,Ali-MNN,chen2018tvm,niu2020patdnn,ma2019pconv} that only only support dense models or pattern-based pruned models, our compiler optimizations are general, support both dense (unpruned) model and sparse (pruned) model with different pruning schemes for fast inference on various mobile platforms. The optimizations include 1) a layer fusion mechanism to fuse different layers together for the reduction of memory consumption of intermediate results and number of operators; 2) the supports for sparse models with different pruning schemes; 3) an auto-tuning process to determine the best-suited configurations of parameters for different mobile CPUs/GPUs; 4) Domain Specific Language (DSL) based code generation. 

\subsection{Layer Fusion Mechanism}
To reduce the inference latency effectively for dense (unpruned) models, a layer fusion technique is incorporated in our compiler optimization to fuse the computation operators in the computation graph. With layer fusion, both the memory consumption of the intermediate results and the number of operators can be reduced. The fusion candidates in a model are identified based on two kinds of polynomial calculation properties, i.e., compression laws and data access patterns. The compression laws include associative property, communicative property, and distributive property. 

However, looking for the fusion candidates in such a large space of all combinations of computation operations is too expensive. Therefore, we introduce two constraints to guide the looking up process: 1) only explore the opportunities that are specifically provided due to the above properties, and 2) only consider enlarging the overall computation for CPU/GPU utilization improvement and reducing the memory access for memory performance improvement as the cost metrics in the fusion. Compared with prior works on loop fusion \cite{ashari2015optimizing,bezanson2017julia,boehm2018optimizing}, our method is more aggressive without high exploration cost.

\subsection{Supports for Different Pruning Schemes}
\begin{figure} [t]
     \centering
     \includegraphics[width=1\columnwidth]{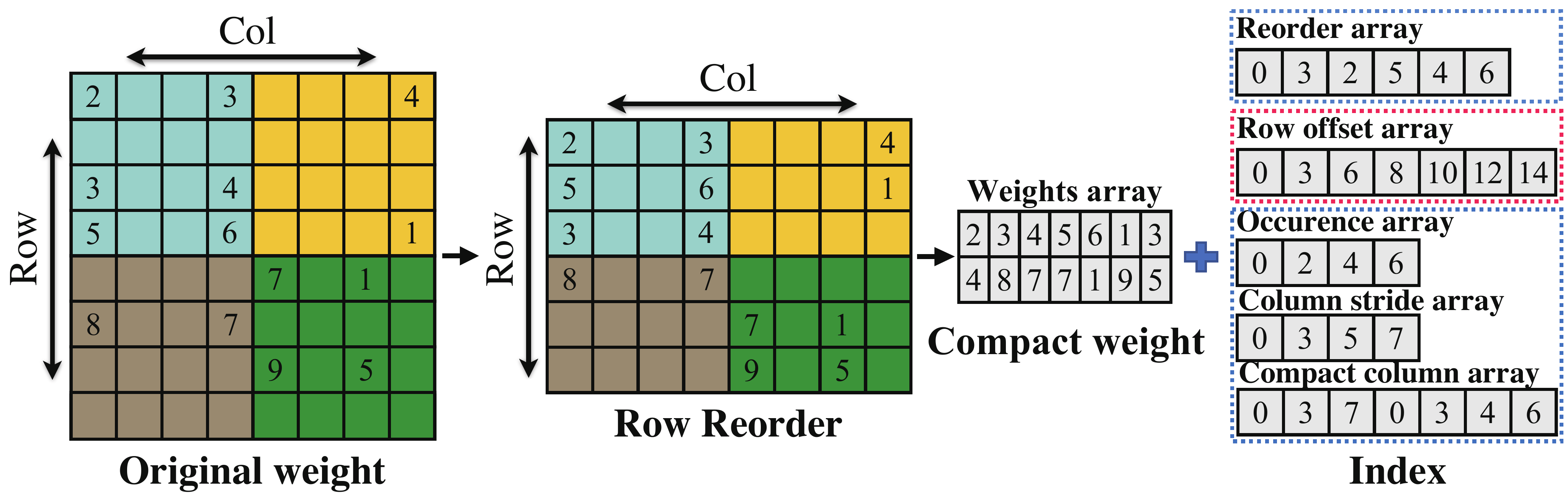}     
     \caption{Matrix Reorder and Blocked Compressed Storage (BCS) for weights.}
     \label{fig:compiler_block}
\end{figure}

Different from other DNN inference frameworks, our framework also supports sparse model accelerations with different pruning schemes including unstructured pruning, coarse-grained structured pruning, pattern-based pruning, and block-based pruning. Note that fine-grained unstructured pruning and coarse-grained structured pruning can be viewed as two extreme cases of block-based pruning by adopting $1\times1$ and the whole weight matrix size as the block size, respectively. Thus, accelerating block-based pruned model also indicates the inference speedup for unstructured pruned and the traditional structured pruned models. 
For the sparse (pruned) model, the framework first compacts the model storage with a novel compression format called Blocked Compressed Storage (\textbf{BCS}) format, as shown in Figure~\ref{fig:compiler_block}, and then performs computation reordering to reduce the branches within each thread and eliminate the load imbalance among threads.



\textbf{BCS} stores non-zero weights as Compressed Sparse Row format (CSR) with an even better compression rate by further compressing the index with a hierarchical structure. Traditional CSR has to store each non-zero weight with an explicit column index. Take block-based pruning as an example, it preserves non-zero weights in identical columns in each block, inducing many repeated column indices if we use CSR. BCS eliminates this redundancy with a hierarchical compression on the column index only. 

For block-based pruning, it partitions the weight matrix of a whole layer into blocks with different pruning configurations. 
Without any further optimization, it will encounter the well-known challenges for sparse matrix multiplications, i.e., heavy control-flows within each thread, load imbalance among multiple threads, and irregular memory access. To address this issue, a row reordering optimization is also included to further improve the regularity of the weight matrix. After this reordering, the continuous rows with identical or similar numbers of non-zero weights are processed by multi-threads simultaneously, thus eliminating thread divergence and achieving load balance.

Figure~\ref{fig:compiler_block} shows a simplified example. {\tt Weights array} stores all non-zero weights. {\tt Compact column array} stores the compressed column index, e.g., [0, 3, 7] denotes the column id of the first three weights [2, 3, 4]. {\tt Column stride array} denotes the starting and ending index of each row in compact column array, e.g., [0, 3] denotes that the column index for the first row starts from index 0 and ends at index 2 in compact column array. The same column indices may be used for multiple rows. {\tt Occurrence array} is used to specify the starting and ending rows with the identical column index, e.g., [0, 2] means that row 0 and 1 share the same column index. BCS also contains a {\tt row offset array} to specify the starting location of each row in weight array.

Usually, the weight distribution is not as regular as the above simplified example, thus, a row reordering optimization is also included to further improve the regularity of the weight matrix. After this reordering, the continuous rows with identical or similar numbers of non-zero weights are processed by multi-threads simultaneously, thus eliminating thread divergence and achieving load balance. Each thread processes more than one rows, thus eliminating branches and improving instruction-level parallelism.   
Moreover, a similar optimization flow (i.e., model compaction and computation reorder and other optimizations) is employed to support all compiler optimizations for pattern-based pruning as PatDNN~\cite{niu2020patdnn}.

\subsection{Auto-Tuning for Different Mobile CPUs/GPUs}

During DNN execution, there are many tuning parameters, e.g., matrix tiling sizes, loop unrolling factors, and data placement on GPU memory, that influence the performance. It is hard to determine the best-suited configuration of these parameters manually. To alleviate this problem, our compiler incorporates an auto-tuning approach for both sparse (pruned) model and dense (unpruned) model. The Genetic Algorithm is leveraged to explore the best-suited configurations automatically. It starts parameter search after an initialization with an arbitrary number of chromosomes and explores the parallelism better. Acceleration codes for different DNN models and different mobile CPUs/GPUs can be generated efficiently and quickly through this  auto-tuning process, providing the foundation for fast end-to-end inference. The auto-tuning optimizations together with the layer-fusion optimizations make our framework outperform other acceleration frameworks. 

\subsection{DSL based Code Generation}
In deep learning, a computational graph of a DNN model can be represented by a directed acyclic graph (DAG). Each node in this graph corresponds to an operator.
We propose a high-level Domain Specific Language (DSL) to specify such kind of operators. Each operator in a computational graph also with a layerwise Intermediate Representation (IR) which contains BCS pruning information. The input and output are different tensors in terms of different shapes. This DSL also provides a {\tt Tensor} function for users to create matrices (or tensors).

In this way, DSL is equivalent to a computational graph (that is, DSL is another type of high-level functions used to simulate the data flow of the DNN model), and they can be easily converted to each other. DSL provides users with the flexibility to use existing DNNs or create new DNNs, improving the productivity of DNN programming.
If the DNN already exists, we will convert it into an optimized calculation graph and convert this graph into a DSL. Otherwise, the user writes the model code in the DSL, converts it back to a calculation graph, performs advanced optimization, and regenerates the optimized DSL code.

Finally, our compiler translates the DSL into low-level C++ code for mobile CPU and OpenCL code for mobile GPU, and optimizes the low-level code through a set of optimizations enabled by BCS pruning. The generated code can be then deployed on the mobile device.

\section{Reweighted $\ell_{1}$ Algorithm} \label{app: reweightedl1}
Prior weight pruning algorithms such as using the group Lasso regularization~\cite{wen2016learning,he2017channel,liu2017learning} or Alternating Direction Methods of Multipliers (ADMM)~\cite{zhang2018systematic,ren2019admm,li2019compressing} either suffer from potential accuracy loss or require manually compression rate tuning.
To overcome the limitations, we leverage reweighted group Lasso~\cite{candes2008enhancing} method to discover the sparsity with automatically determined pruning ratio. The basic idea is to systematically and dynamically reweight the penalties. More specifically, the reweighted method reduces the penalties on weights with larger magnitudes, which are likely to be more critical weights, and increases the penalties on weights with smaller magnitudes.

We formulate the general reweighted pruning problem as below
{\small
\begin{equation} \label{prob_block}
 \underset{\bm{W}}{\text{minimize\quad}}
 f \big( \bm{W};\mathcal{D} \big)+\lambda\sum_{l=1}^{L} R(\bm{\alpha}^{l},\bm{W}^{l}), \\
\end{equation}}%
where $\lambda$ is the hyperparameter to adjust the relative importance between accuracy and sparsity. $\mathcal{D} $  stands for the dataset. $\bm{W}^l$ denotes the weight matrix for the $l$-th layer and $\bm{W} := \{\bm{W}^l\}_{l=1}^L$. 
Let $\bm{\alpha}^{l}$ represent the collection of penalty values that applied on the  $l$-th layer weights. Note that each element in $\bm{\alpha}^{l}$ is  positive.

The regularization term  $R(\bm{\alpha}^{l},\bm{W}^{l})$  denotes the penalties on the weights  for the $l$-th layer. The method can be applied to models with different pruning schemes for each layer. For block-based pruning, each $\bm{W}^{l}$ is divided into $J$ blocks with the same size $p^l\times q^l$, namely, $\bm W^l = [\bm W_{1}^l,  \bm W_{2}^l,...,\bm W_{J}^l]$, and the regularization term for block-based column pruning is defined as
\begin{equation}
\small
R({\bm{\alpha}}^{l},\bm{W}^{l}) = \sum_{j=1}^{J}\sum_{n=1}^{q^l} \big\| \alpha_{jn}^{l} \cdot  [\bm W_{j}^{l}]_{:,n} \big\|_F^2,
\label{eq:column}
\end{equation}
where $[\bm W_{j}^l]_{:,n}$ is the $n$-th column of $\bm W_{j}^l$ and $\alpha^{l}_{jn}$ is updated by $\frac{1}{\|[\bm W_{j}^{l}]_{:,n}\|_F^2 + \epsilon}$ to help increase the degree of sparsity.  $ \epsilon$ is a small value to avoid zero denominator. From the equation we can see that small $\| [\bm W_{j}^{l}]_{:,n}\|_F^2$ leads to a large penalty $\alpha_{jn}^l$, thus is more likely to be pruned. 

Similarly, the regularization term for block-based row pruning is defined as
\begin{equation}
\small
R({\bm{\alpha}}^{l},\bm{W}^{l}) = \sum_{j=1}^{J}\sum_{m=1}^{p^l} \big\| \alpha_{jm}^{l} \cdot  [\bm W_{j}^{l}]_{m,:} \big\|_F^2,
\label{eq:row}
\end{equation}
where $[\bm W_{j}^{l}]_{m,:}$ represents the $m$-th row of $\bm W_{j}^l$ and $\alpha^{l}_{jm}$ is updated by $\frac{1}{\|[\bm W_{j}^{l}]_{m,:}\|_F^2 + \epsilon}$.
For coarse-grained structured pruning, it can be viewed as the special case of block-based pruning by setting $J=1$ and $p^l\times q^l$ as the original weight matrix size of layer $l$ in equation (\ref{eq:column}) and equation (\ref{eq:row}).

As for pattern-based pruning, as it acts on the kernel levels and suits tensor-based computation better, we formulate it with tensor representations. We represent the weight tensor for the $l$-th layer as $\mathcal{W}^l \in \mathbb{R}^{N^l\times C^l \times K_h^l \times K_w^l}$, where $N^l$, $C^l$, $K_h^l$, $K_w^l$ represent the number of filters, the number of channels, kernel height and kernel width for the $l$-th layer, respectively. We first apply a kernel pattern from a pre-defined kernel pattern library to each $3\times3$ kernel in the model, resulting in weight tensor $\mathcal{W'}^l$ for $l=1,\cdots,L$. We further apply connectivity pruning and the regularization term is defined as
\begin{equation}
\small
R({\bm{\alpha}}^{l},\mathcal{W'}^{l}) = \sum_{n=1}^{N^l}\sum_{m=1}^{C^l} \big\| \alpha_{nm}^{l} \cdot  [\mathcal {W'}^{l}]_{n,m,:,:} \big\|_F^2,
\label{eq:connectivity}
\end{equation}
where $[\mathcal {W'}^{l}]_{n,m,:,:}$ stands for the kernel that connects the $m$-th input channel with the $n$-th output channel, and $\alpha^{l}_{nm}$ is updated by $\frac{1}{\|[\mathcal{W'}^{l}]_{n,m,:,:}\|_F^2 + \epsilon}$.

In each iteration of the prune ratio determination, we solve  problem (\ref{prob_block}) with certain epochs of training. Then with the obtained $\bm W$, we can update $\bm \alpha$. Thus in the next iteration, we again solve  problem (\ref{prob_block})  with updated $\bm \alpha$. After iterations, we can obtain the sparse weights without human intervention.
We see that the reweighted method only requires the hyperparameter $\lambda$ and the soft constraints formulation allows the automatic determination of the prune ratio  for each layer. 

\section{Comparison with State-of-the-Art for $\times$3 Scaling Task} \label{sec:3upscaling_result}

\begin{table*}[t]
\scriptsize\sffamily
    \centering
    \renewcommand{\arraystretch}{1.2}
\begin{adjustbox}{max width=1\textwidth}
\begin{tabular}{@{} l l *{6}{r} @{}}
\toprule
\bfseries Scale
& \bfseries \begin{tabular}{@{}l@{}}Model\end{tabular}
& \bfseries \begin{tabular}{@{}r@{}}Params \\(K)\end{tabular}
& \bfseries \begin{tabular}{@{}r@{}}Multi-Adds \\(G)\end{tabular}
& \bfseries \begin{tabular}{@{}r@{}}Set5\\(PSNR/SSIM)\end{tabular}
& \bfseries \begin{tabular}{@{}r@{}}Set14\\(PSNR/SSIM)\end{tabular}
& \bfseries \begin{tabular}{@{}r@{}}B100\\(PSNR/SSIM)\end{tabular}
& \bfseries \begin{tabular}{@{}r@{}}Urban100\\(PSNR/SSIM)\end{tabular}
\\[2ex]
    \toprule
    \multirow{9}{*}{\rotatebox[origin=c]{0}{\footnotesize $\times$ 3}} & \textsc{SRCNN}~\cite{dong2014learning} & 57 & 52.7 & 32.75/0.9090 &29.28/0.8209 &28.41/0.7863 &26.24/0.7989 \\
    & \textsc{FSRCNN}~\cite{dong2016accelerating} & 12& 5.0& 33.16/0.9140& 29.43/0.8242 &28.53/0.7910 &26.43/0.8080 \\
    & \textsc{CARN-M}~\cite{hui2018fast} & 412 &46.1& 33.99/0.9236 &30.08/0.8367 &28.91/0.8000 &27.55/0.8385\\
    & \textsc{ESRN-V}~\cite{song2020efficient} & 324 &36.2 &34.23/0.9262 &30.27/0.8400 &29.03/0.8039 &27.95/0.8481 \\
    & \textsc{EDSR}~\cite{lim2017enhanced} & 1518 & 160.8 & 34.37/0.9270 & 30.28/0.8418 &29.09/0.8052 &28.15/0.8527 \\
    & \textsc{WDSR}~\cite{yu2018wide} & 1203 & 122.5 & 34.48/0.9279 &30.39/0.8434 &29.16/0.8067 &28.38/0.8567 \\
    & \textbf{Ours ($t=290$}ms\textbf{)}  & 122 & 12.5 & 34.13/0.9252 & 30.12/0.8372 & 28.98/0.8015 & 27.71/0.8420 \\
    & \textbf{Ours ($t=150$}ms\textbf{)}  & 51 & 5.2 & 33.85/0.9225 & 29.95/0.8347 & 28.86/0.7984 & 27.35/0.8340  \\
    & \textbf{Ours ($t=50$}ms,\textbf{real-time)}  & 16 & 1.5 &  33.29/0.9160 & 29.57/0.8261 & 28.61/0.7929 & 26.44/0.8106  \\
    \bottomrule
\end{tabular}
\end{adjustbox}
    \caption{Comparison of searched results with state-of-the-art efficient SR models for the $\times$3 upscaling task.}
    \label{app_table:result_sr}
\end{table*}

Besides the results for $\times2$ and $\times4$ upscaling task, we further compare our models on $\times$3 upscaling with state-of-the-art efficient SR models. As shown in Table \ref{app_table:result_sr}, for the $\times$3 upscaling task, our model obtained with a target latency $t=150$ms reaches higher PSNR/SSIM than SRCNN and FSRCNN with comparable or even fewer MACs. With $t=290$ms, our model provides better PSNR/SSIM compared with CARN-M with $3.7\times$ MACs reduction. Compared with ESRN-V, EDSR, and WDSR, competitive image quality in terms of PSNR/SSIM can be obtained by our model with much fewer MACs. By setting $t=50$ms, our model reaches real-time inference while keeping fairly good PSNR/SSIM. 

\section{Visual Comparison with Other SR methods} \label{sec:visual_compare}

\begin{figure}[t]
     \centering
     \includegraphics[width=1.0\columnwidth]{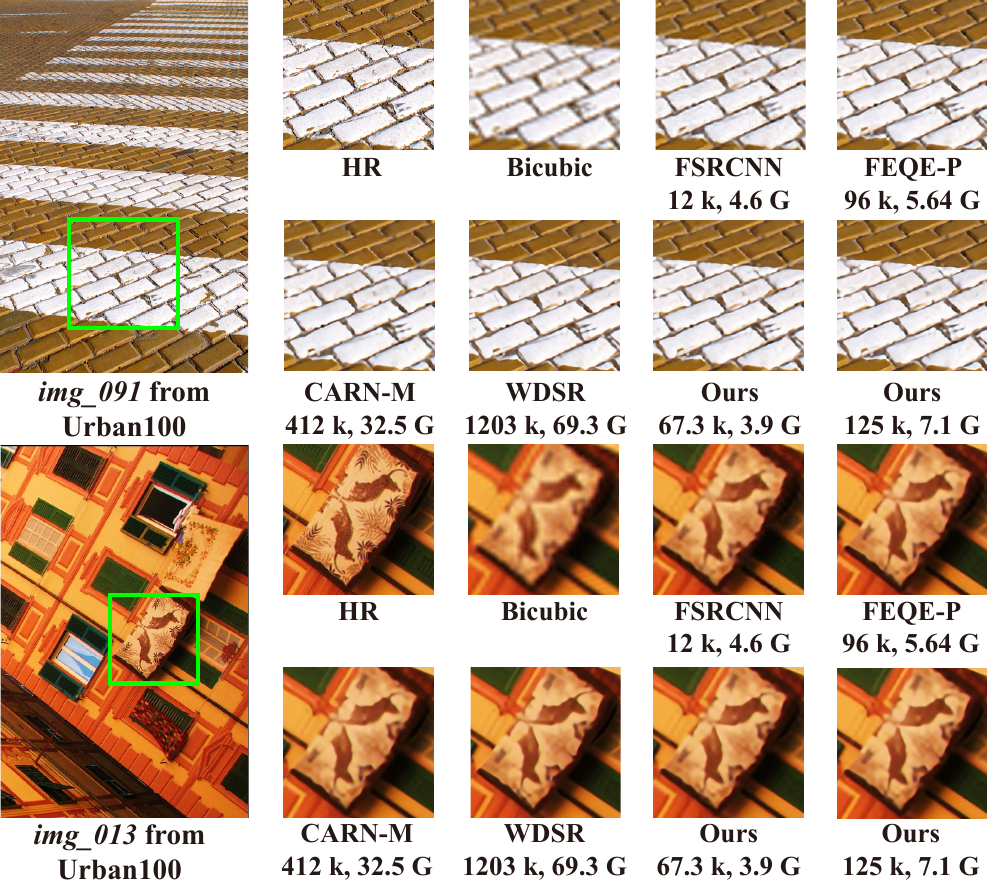}  
     \caption{Visual comparison with other SR models on $\times$4 scale. Model parameters and MACs are listed under model name.}
    \label{fig:patch2}  
\end{figure}

In this section, we include more visual comparisons with other SR models on $\times4$ upscaling task, as shown in Figure \ref{fig:patch2}. The low-resolution images are img\_091 and img\_013 from Urban100. As observed, high-resolution images generated by our models demonstrate undetectable visual difference compared with the baseline WDSR while greatly saving the parameters and MACs. 

\section{Ablation Study} \label{app: mnn}

We compare our method with other SR models in terms of FPS and PSNR. For a fair comparison, we implement our derived models and other baselines with MNN (not support sparse models for further inference accelerations) on mobile CPU as the baselines. We want to promote reproducibility and evaluate speedup using the same framework to show the generalization of our searched results. Note that we modify the sparse models derived by our method by filling each pruned weight with a zero value as MNN does not support sparse models for further speedups. In this way, the models being dealt with MNN are dense models with a bunch of zero-value weights. 

\begin{figure} [t]
     \centering
     \includegraphics[width=1\columnwidth]{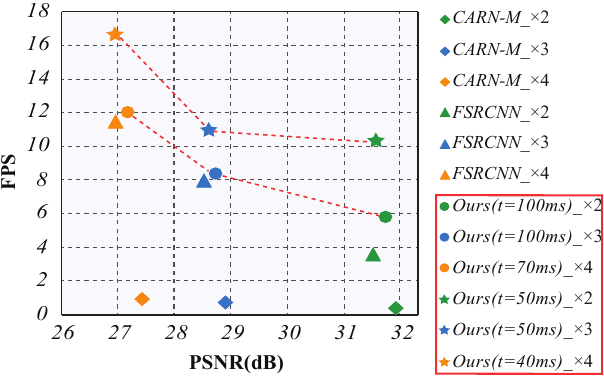}     
     \caption{FPS v.s. PSNR for different SR methods implemented with MNN on mobile CPU evaluated on B100.}
     \label{fig: fps_psnr}
\end{figure}

As shown in Figure \ref{fig: fps_psnr}, compared with CARN-M \cite{hui2018fast} and FSRCNN \cite{dong2016accelerating}, our method with large $t$ can result in higher FPS and PSNR. 
The derived models with smaller $t$ have slight PSNR degradation with   significant FPS improvements.
Note that our model with compiler optimization can satisfy the real-time requirement (such as $t=50$ms). And our derived model implemented with MNN still maintain fairly good results (such as 10.8FPS for the $t=50$ms searched result on the $\times$3 upscaling task) as it demonstrated in Figure \ref{fig: fps_psnr}.

\section{Fast Evaluation for Architecture and Pruning Search}
Though Bayesian Optimization (BO) is leveraged to reduce the evaluation cost, it is still time-consuming to get the precise image quality of each candidate $g$ in the selected $B$ candidates as it requires the complex pruning and the full retraining process. Conducting the architecture and pruning search process with such an evaluation method will take a huge amount of time and computations. To reduce the overall search time, we utilize several strategies to accelerate the image quality evaluation process. 

First, instead of using a complex pruning algorithm such as iterative pruning \cite{han2015} and regularization-based methods \cite{ren2019admm,zhang2018systematic}, we conduct a magnitude-based one-shot pruning by removing weights based on the L2-norm of the selected structural sparsity and the pruning ratio according to the latency model. Though it might lead to a more severe accuracy degradation compared with other pruning methods, one-shot pruning can still distinguish the different performance among different pruning schemes. Moreover, it is the relative accuracy performance, not the precise accuracy of different pruning schemes, that the search process cares for. Therefore, adopting a magnitude-based one-shot pruning for fast evaluation if suitable.

Second, as the supernet is well-trained with each candidate dense net $a \in \mathcal{A}$ is optimized simultaneously, we leverage an early stopping mechanism in the retraining phase by only retraining for several epochs. The partially regained accuracy can predict the final model accuracy and be used to compare the performance among different schemes~\cite{zhong2018practical, tan2019mnasnet}. With the fast pruning and retraining of each selected candidate $g$, we could greatly accelerate the image quality evaluation, thus significantly reducing the search cost. 

\section{LPIPS Performance}
We further evaluate the perceptual quality of our method in terms of  LPIPS. The results are shown in Table~\ref{table:yolov4}. According to the results, our method needs much less resource with second-best LPIPS.

\begin{table}[!t]
\setlength\tabcolsep{2pt} 
\footnotesize
\centering
\scalebox{0.95}{
\begin{tabular}{c|c|c|c|c|c|c|c}
\toprule

Metric & Method & Params & Multi-Adds & Set5 & Set14 & B100 & Urban100  \\ \hline
&FSRCNN & 12K & 4.6G& 0.2187&0.3032&0.3354&0.3414\\ 
&CARN-M &412K &32.5G& 0.1810 &0.2733 &0.3128&0.2793 \\ 
LPIPS&WDSR & 1203K &69.3G& \textcolor{red}{0.1764} &\textcolor{red}{0.2640}&\textcolor{red}{0.3047}&\textcolor{red}{0.2535} \\ 
&Ours &125K&7.1G & \textcolor{blue}{0.1793}&\textcolor{blue}{0.2725}&\textcolor{blue}{0.3117}&\textcolor{blue}{0.2742} \\ 
&Ours & 12K & 0.7G & 0.1954&0.2882&0.3218&0.3135\\ \bottomrule
\end{tabular}}
\caption{Comparison on ×4 upscaling tasks using LPIPS. Lower is better for LPIPS. Multi-Adds is reported for an input 320×180 image patch. Red/blue
text: best/second-best LPIPS result.}
\label{table:yolov4}
\vspace{-12pt}
\end{table}

\section{Comparison with APQ}
APQ~\cite{wang2020apq} jointly searches network architecture, pruning, and quantization for efficient DNN deployment. The differences between our method and APQ are summarized as below: (i) We have different search objectives. While APQ focuses on classification accuracy, we try to achieve real-time (RT) Super Resolution (SR) on mobile with specific RT requirements, \textcolor{black}{which is more challenging due to  huge computations for high resolutions}. 
(ii) We have different search strategies.  We decouple   pruning ratio search from   architecture and pruning scheme search to reduce  search complexity, \textcolor{black}{thus accelerating the search process}, while APQ   unifies NAS, pruning and quantization as joint optimization. (iii) We have a larger pruning scheme search space. We can choose channel pruning, pattern  pruning, or block  pruning  for each layer with higher flexibility \textcolor{black}{for both high accuracy and speed}, while APQ only supports coarse-grained channel pruning with potential accuracy degradation when  pruning ratio is high.  (iv) We adopt \textcolor{black}{ additional} methods (Bayesian optimization with neural predictors) to reduce search cost with higher efficiency. 
(v) The connection with hardware is different. 
We have a specific target hardware device (mobile phones) with detailed compiler optimization (CO), while APQ uses an ASIC design and optimizes model for it.
\end{document}